\documentclass[conferences]{IEEEtran}

\usepackage{floatflt}
\usepackage{color}
\usepackage{cite}
\usepackage{graphicx}
\usepackage{amsmath}
\usepackage{amsthm}
\usepackage{amssymb}
\usepackage{verbatim}
\usepackage{psfrag,array}
\usepackage{subfigure}
\usepackage{multirow}
\usepackage{hhline}
\usepackage{stfloats}
\usepackage{amsmath}
\usepackage{epstopdf}
\usepackage{algorithm}
\usepackage{algpseudocode}
\usepackage{multirow}

\newcommand{\p}[1]{\mathop{\mbox{\it p} } }

\renewcommand{\vec}[1]{\ensuremath{\boldsymbol{#1}}}
\newcommand{\be}{\begin{equation}}
\newcommand{\ee}{\end{equation}}
\newcommand{\ba}{\begin{array}}
\newcommand{\ea}{\end{array}}
\newcommand{\bea}{\begin{eqnarray}}
\newcommand{\eea}{\end{eqnarray}}
\newcommand{\bean}{\begin{eqnarray*}}
\newcommand{\eean}{\end{eqnarray*}}
\newcommand{\argmax}{\mathop{\arg\max}}
\newcommand{\argmin}{\mathop{\arg\min}}

\newcommand{\Ts}{T_\mathrm{s}}

\definecolor{white}{rgb}{1,1,1}

\newtheorem{example}{Example}

\makeatletter

\newcommand*{\algrule}[1][\algorithmicindent]{\makebox[#1][l]{\hspace*{.5em}\vrule height .75\baselineskip depth .25\baselineskip}}%

\newcount\ALG@printindent@tempcnta
\def\ALG@printindent{%
    \ifnum \theALG@nested>0
    \ifx\ALG@text\ALG@x@notext
    \addvspace{-3pt}
    \else
    \unskip
    \ALG@printindent@tempcnta=1
    \loop
    \algrule[\csname ALG@ind@\the\ALG@printindent@tempcnta\endcsname]%
    \advance \ALG@printindent@tempcnta 1
    \ifnum \ALG@printindent@tempcnta<\numexpr\theALG@nested+1\relax%
    \repeat
    \fi
    \fi
}%
\usepackage{etoolbox}
\patchcmd{\ALG@doentity}{\noindent\hskip\ALG@tlm}{\ALG@printindent}{}{\errmessage{failed to patch}}
\makeatother
\makeatother

\begin{document}

\title{On Time-of-Arrival Estimation in NB-IoT Systems}
\author
{
Sha Hu, Xuhong Li, and Fredrik Rusek  \\
Department of Electrical and Information Technology, Lund University, Lund, Sweden \\
\{firstname.lastname\}@eit.lth.se
}

\maketitle

\begin{abstract}
We consider time-of-arrival (ToA) estimation of a first arrival-path for a device working in narrowband Internet-of-Things (NB-IoT) systems. Due to a limited 180 KHz bandwidth used in NB-IoT, the time-domain auto-correlation function (ACF) of transmitted NB positioning reference signal (NPRS) has a wide main lobe. Without considering that, the performance of ToA estimation can be degraded for two reasons. Firstly, under multiple-path channel environments, the NPRS corresponding to different received paths are superimposed on each other, and so are the cross-correlations corresponding to them. Secondly, the measured peak-to-average-power-ratio (PAPR) used for detecting the presence of NPRS is inaccurate. Therefore, in this paper we propose a space-alternating generalized expectation-maximization (SAGE) based method to jointly estimate the number of channel taps, the channel coefficients and the corresponding delays in NB-IoT systems, with considering the imperfect ACF of NPRS. Such a proposed method only uses the time-domain cross-correlations between the received signal and the transmitted NPRS, and has a low complexity. We show through simulations that, the ToA estimation of the proposed method performs close to the maximum likelihood (ML) estimation for a single-path channel, and significantly outperforms a traditional ToA estimator that uses signal-to-noise (SNR) or power thresholds based estimation.

\end{abstract}

\section{Introduction}
Observed-time-difference-of-arrival (OTDOA) is a downlink positioning method adopted in long-term-evolution (LTE) \cite{3GPP}, which relies on time-of-arrival (ToA) estimates from at least three base-stations. The 3rd Generation Partnership Project (3GPP) has dedicated a significant effort to enhance positioning support for newly featured narrowband Internet-of-Things (NB-IoT) systems \cite{3GPP, R1}. The NB positioning-reference-signal (NPRS) is transmitted to enhance positioning measurements at receivers to ensure sufficiently high signal quality and detection probability, which is distributed in time and frequency resources over a subframe, and a number of consecutive positioning subframes can be allocated with a certain periodicity. In a subframe where NPRS is present, no data but only control signalings are transmitted to reduce interference such as depicted in Fig. 1.

In NB-IoT systems, the quadrature phase-shift keying (QPSK) modulated NPRS symbols are generated and mapped to one physical resource block (PRB) (180 KHz)~\cite{3GPP}. A number of consecutive NPRS subframes can be configured and transmitted periodically in every radio frame (10ms), and the period of one positioning occasion can be configured to more than a second. The minimum unit of ToA used in LTE is $\Ts\!=\!1/30.72$ $\mu$s, which corresponds to a distance-resolution about 10 meters (m). To successfully position an NB-IoT device, ToA for at least three base-stations need to be detected. With more detected base-stations, the positioning resolution can be further improved. 

\begin{figure}[t]
\begin{center}
\vspace*{1mm}
\hspace*{-0mm}
\scalebox{1.25}{\includegraphics{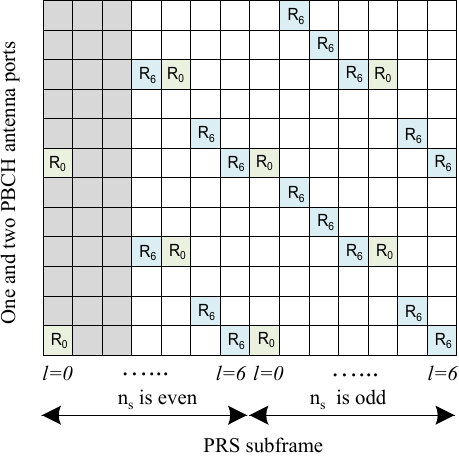}}
\vspace*{-2mm}
\caption{\label{fig2}An NPRS pattern in one PRB with normal cyclic-prefix (CP) in LTE. The NPRS transmitted on antenna port 6 is labeled as $R_6$, while the normal cell-specific reference signal (CRS) sent on antenna port 0 is labeled as $R_0$.}
\vspace*{-7mm}
\end{center}
\end{figure}

A rich literature exists for ToA estimation which can be considered as channel estimation problem, and investigations have been carried out for ultra wideband (UWB) and OFDM systems such as in \cite{VGM10, PJ09}. For NB-IoT system, ToA estimation becomes more challenging due to a limited number of NPRS symbols transmitted in a bandwidth 180 KHz, even though the channels are assumed constant over one subframe due to low mobility of NB-IoT devices. As shown in \cite{TT92}, it is difficult to distinguish two arriving paths with a time-delay difference less than an inverse of the signal bandwidth \cite{TT92}. Under multiple-path environments, the imperfect NPRS auto-correlation functions (ACF) corresponding to different received paths are superimposed on each other, and degrades the ToA estimation performance. Further, even under a single-path channel such as AWGN, the measured peak-to-average-power-ratio (PAPR) used for detecting ToA is inaccurate without considering the wide lobe of the ACF. 

Subspace fitting algorithms such as multiple signal classification (MUSIC) \cite{LP04} and estimation of signal parameters via rotational invariance techniques (ESPRIT) \cite{PJ09} can be efficient in estimating the ToA with superimposed signals. But the computational complexities are relative high and Fourier transformations are needed to transfer the received signal into frequency domain. In \cite{VGM10}, the authors propose a low-complexity extension of the sliding correlator for channel and ToA estimations based on maximum likelihood (ML) principle, in the presence of pulse overlap for impulse-radio UWB system. However, the authors assume that, the number of channel taps are known and no overlap between three or more consecutive paths, which may not apply in NB-IoT systems. Traditionally, there are also threshold-based algorithms for ToA estimation utilizing such as measured signal-to-noise (SNR) and power \cite{LS14, XHZD14, Ret15}. In \cite{RZS16}, the authors propose an iterative algorithm for ToA estimation based on the ACF, which iteratively cancels the peaks from the obtained cross-correlations. However, as the peaks are incorrect caused by overlapping of ACFs corresponding to different channel taps, it does not improve the performance of ToA estimation for NB-IoT systems \cite{HBR17}.

In this paper, we consider ToA estimation in NB-IoT systems where only one PRB is used for data-transmission. As the ML function depends on the unknown number of channel taps, the channel coefficients and the delays, the optimal estimation has prohibitive complexity. Therefore, we propose a space-alternating generalized expectation-maximization (SAGE) based method to jointly estimate these parameters and the ToA, with considering the ACF of the time-domain NPRS signal. The proposed algorithm is based on the time-domain cross-correlations of the received signal and the transmitted NPRS signal which yields a low computational cost, and mainly comprises two steps running iteratively: The first step uses the SAGE algorithm to estimate the channel coefficients and the corresponding delays for a given number of channel taps; While the second step uses heuristic approaches to remove the invalid channel taps. We show through simulations results that, the proposed ToA estimation method works well both for a single-path channel such as AWGN and multi-path fading channels, which is difficult for traditional threshold based ToA estimators to achieve.

\section{Problem Formulation}

\subsection{Received Signal Model}
In the considered single-input-single-output (SISO) NB-IoT system, we assume a sampling-rate $\tilde{F}_{ s}$ with a typical value 1.92 MHz. Denote the true delay of the $i$th channel tap as $\tau_{i}$, which is measured in number of samples as
\bea  d_{i}=\lfloor\tau_{i}\tilde{F}_{s}\rfloor. \eea
Then, the superimposed received samples $y[n]$ with a total length $D$ for all $L$ taps can be modeled as
\bea \label{md1}  y[n]=\sum_{i=0}^{L-1}h_{i}s[n\!-\!d_{i}]+w[n],\eea
where $s[n]$ is the time-domain NPRS signal (including CP) transmitted in one subframe with a total length $S$, and $h_{i}$ are complex-valued channel coefficients that are assumed constant over one subframe. The noise $w[n]$ is modeled as AWGN with a zero-mean and variance $\sigma^2$, and the SNR is defined as \bea \mathrm{SNR}\!=\!\sigma_{\mathrm{s}}^2/\sigma^2,\eea
where $\sigma_{\mathrm{s}}^2$ is the averaged power of transmitted $s[n]$. 

Traditionally, in order to detect the ToA, i.e., $d_0$, a cross-correlation between the received samples $y[n]$ and the NPRS $s_[n]$ is implemented with a correlator, which yields
\bea \label{R1} R[d]=\!\sum_{k=d}^{d+S-1}\!\!y[k]s^{\ast}[k-d]=\sum_{i=0}^{L-1}h_i \gamma(d-d_i)+\tilde{w}[d],\eea
where $\gamma(d)$ is the normalized ACF of $s[n]$ with delay $d$, and $\tilde{w}[d]$ is the noise term after correlation which is colored. An NPRS signal is claimed to exist if a following condition
\bea \label{threshold1} \mathrm{PAPR}\triangleq \frac{\underset{d}{\max} \left\{|R[d]|\right\}}{\frac{1}{D}\sum\limits_{d=0}^{D\!-\!1}\!|R[d]|}\!>\!\eta_1,\eea
is satisfied, where $D$ is the length of $R[d]$ that corresponds to a possible maximum time-delay. With (\ref{threshold1}) satisfied, a threshold based method \cite{Ret15, VGM10} estimates the ToA $d_{0}$ according to
\bea \label{threshold2} d_{0}=\underset{d}{\mathrm{argmin\;\;}} \left\{ \frac{|R[d]|}{\underset{d}{\max} \left\{|R[d]|\right\}}>\eta_2 \right\}\!. \eea
In (\ref{threshold1}) and (\ref{threshold2}), $\eta_1$ and $\eta_2$ are predefined thresholds that can be adjusted for different scenarios.

\begin{figure}[t]
\begin{center}
\scalebox{0.31}{\includegraphics{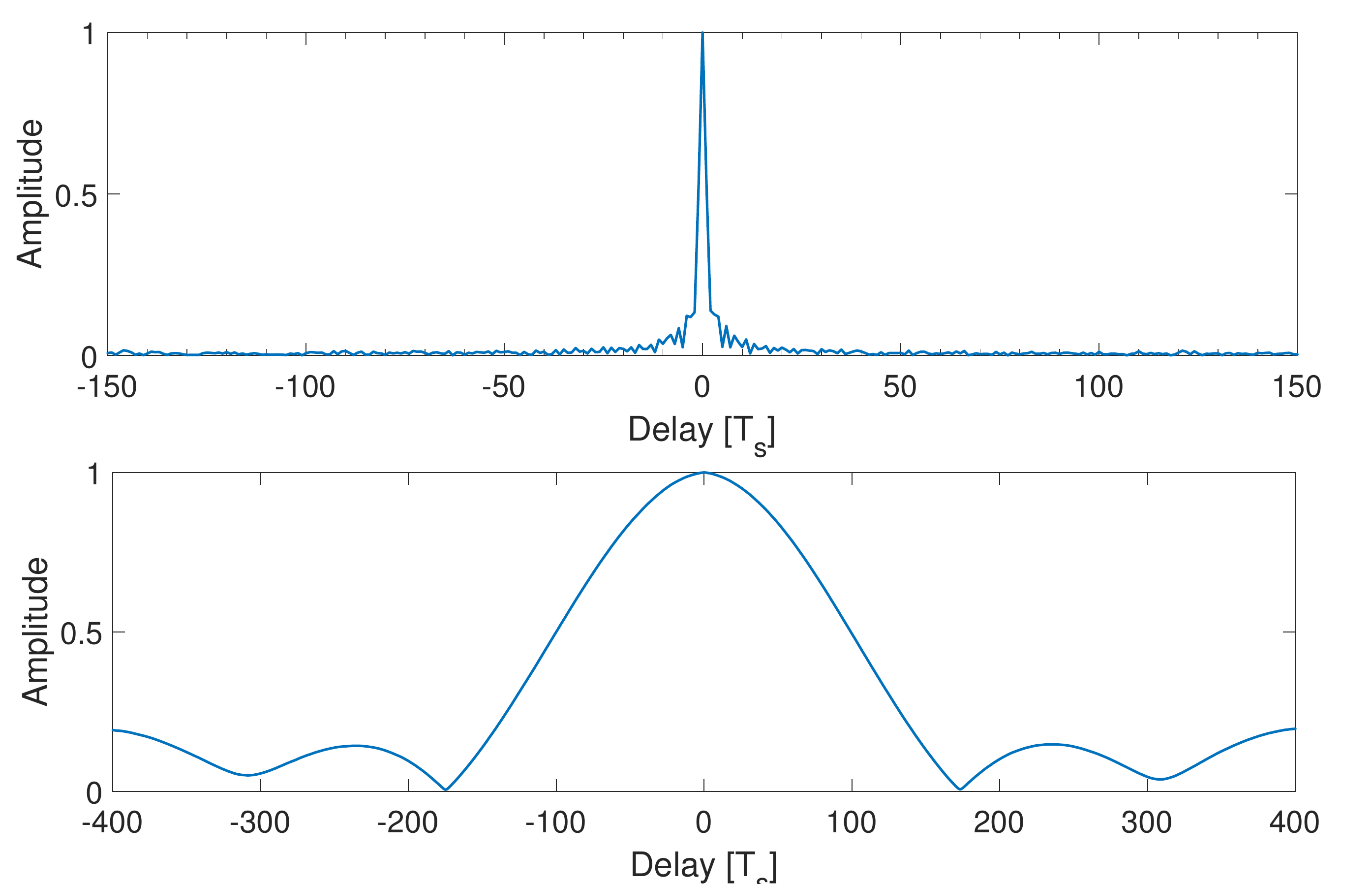}}
\vspace*{-7mm}
\caption{\label{fig2}The amplitude of time-domain ACFs of PRS in one subframe with 100 PRB  and 1 PRB at sampling rate 30.72 MHz, respectively.}
\vspace*{-7mm}
\end{center}
\end{figure}

There are several drawbacks of estimating ToA applying (\ref{threshold1}) and (\ref{threshold2}) in NB-IoT systems. Firstly, when channel coefficients $h_i$ are unkown, the non-coherent addition in (\ref{threshold1}) of $R[d]$ is inferior to a coherent addition that compensates the channel impacts \cite{HBR17}. Secondly, since the ACF of the NPRS has a wide lobe in NB-IoT systems, it should be considered in the PAPR evaluation in (\ref{threshold1}). Thirdly, the estimate $d_0$ in (\ref{threshold2}) can be inaccurate under the case that $\underset{d}{\max} \left\{|R[d]|\right\}$ is attained with the correct ToA such as under a sing-path channel channel. Lastly, the threshold in (\ref{threshold2}) needs to be tuned for different channels types, which is however, unknown to the receivers.

In Fig. 2 the amplitude of normalized time-domain ACF of the NPRS transmitted in 100 PRBs and only one PRB at sampling rate 30.72 MHz are shown. As what can be clearly seen that, with only one PRB used, the ACF has a wide lobe. Hence, the ACFs corresponding to different channel taps in $R[d]$ can be superimposed with each other. We illustrate such an phenomenon through an example below.

\begin{figure}
\begin{center}
\scalebox{0.3}{\includegraphics{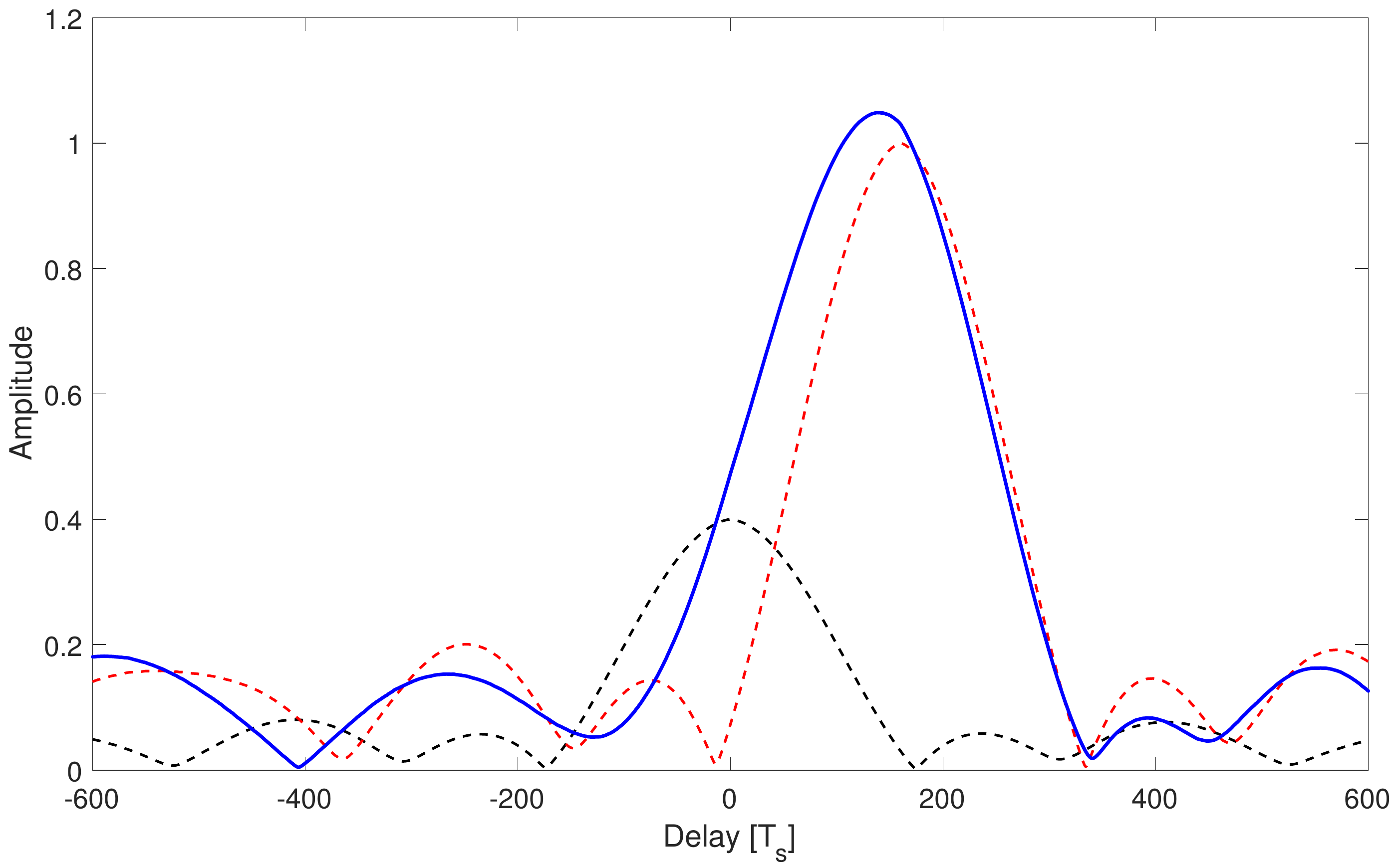}}
\vspace*{-4mm}
\caption{\label{fig3}The superimposed cross-correlations in Example 1, with the correct ToA at 0, but the maximum of $|R(d)|$ is achieved at $d\!=\!138$, due to the strong impact from a second path.}
\vspace*{-7mm}
\end{center}
\end{figure}

\begin{example}
Consider a customized two-path channel with constant coefficients $\vec{h}\!=\![0.4, 1]$ and delays $\vec{\tau}\!=\![0, 160T_s]$. The noiseless cross-correlation $R[d]$ is shown in Fig. \ref{fig3}.
\end{example}

\subsection{Formulation of ToA Estimation}
 
Denote the delay vector $\vec{d}\!=\![d_0,\,d_1,\,\cdots,\,d_{L-1}]$ and the channel vector $\vec{h}\!=\![h_0,\,h_1,\,\cdots,\,h_{L-1}]$. From (\ref{md1}), the ML estimator is
\bea \label{prbm} (\tilde{\vec{d}},\tilde{\vec{h}},\tilde{L})=\argmin_{\vec{d},\vec{h},L} \sum_{n=0}^{N-1}\left|y[n]-\sum_{i=0}^{L-1}h_{i}s[n\!-\!d_{i}]\right|^2 \!, \eea
Directly optimizing (\ref{prbm}) for all possible values of the unknown parameters can yield a prohibitive complexity. Therefore, we next solve (\ref{prbm}) in a suboptimal manner with an objective of good performance for all types of channel and a low computational cost.

\section{Proposed ToA Estimation Method}

The proposed ToA estimation method to solve (\ref{prbm}) mainly comprises two steps. In a first step, we assume $L$ is known and estimate $\vec{h}$ and $\vec{n}$ using SAGE algorithm \cite{H14}. Then, in a second step, heuristic approaches are used to refine $L$ and removing invalid channel taps. These two steps are running iteratively until $L$ converges. As shown in Table 1, an initial value of $L$ can be set to 9, 5 and 2 for $R[d]$ evaluated at 30.72 MHz, 1.92 MHz and 240 KHz, respectively, which is sufficiently large to cope with the largest delays for the listed fading channels.

\subsection{Step 1: SAGE Iteration with a Given $L$}
We first present the SAGE based algorithm which is summarized in Algorithm 1.  If there is no priori information about the channel delays and coefficients, we can sort $|R[d]|$ and set an initial $\tilde{\vec{d}}$ to the locations of the $L$ largest values, and an initial channel estimate as
\bea \tilde{h}_\ell\!=\!R[\tilde{d}_\ell],  \qquad 0\!\leq\!\ell\!<\!L. \eea
Let $\tilde{\vec{d}}$ and $\tilde{\vec{h}}$ be the estimated delays and channel coefficients in a previous iteration. Then, in a next SAGE iteration, the delay of the $\ell$-th tap can be updated as
\bea \label{nl} \tilde{d}_\ell=\argmax_{d}\tilde{R}_{\ell}[d], \eea
where $\tilde{R}_{\ell}[d]$ is the cross-correlation that removes the other estimated NPRS, which equals
\bea \label{tR} \label{R} \tilde{R}_{\ell}[d]=\!\sum_{k=d}^{d+S-1}\!\!\tilde{y}_{\ell}[k]s^{\ast}[k-d].\eea
with
\bea \label{ty} \tilde{y}_{\ell}[n]=y[n]-\sum_{i=0,i\neq\ell}^{L-1}\tilde{h}_{i}s[n\!-\!\tilde{d}_{i}].\eea
Inserting (\ref{ty}) into (\ref{tR}) yields
\bea \label{tR1} \tilde{R}_{\ell}[d]=R[d]-\sum_{i=0,i\neq\ell}^{L-1}\tilde{h}_i \gamma(d-\tilde{d}_i). \eea

With (\ref{tR1}), the noise power is estimated as
\bea \tilde{\sigma}^2=\frac{1}{D}\sum_{d=0}^{D-1}|\tilde{R}_{\ell}[d]-\tilde{h}_\ell \gamma(d-\tilde{d}_\ell)|^2. \eea
Then, $\tilde{d}_\ell$ is updated as in (\ref{nl}), and finally the channel coefficient $\tilde{h}_\ell$ is updated as
\bea \label{th} \tilde{h}_\ell=\frac{1}{2E}\sum_{k=-E}^{E}\gamma^{\ast}(k)\tilde{R}_{\ell}[\tilde{d}_\ell+k], \eea
followed by an LMMSE filtering
\bea \label{th1} \tilde{h}_\ell=\frac{\tilde{h}_\ell}{1+\tilde{\sigma}^2/|\tilde{h}_\ell|^2}. \eea
The variable $E$ is yet to be optimized, which depends on the property of the ACF shown in Fig. 1. In principle, for a single-path channel, it is beneficial to set $E$ to a large value to denoise the estimate $\tilde{h}_\ell$. However, for multiple-path channel, a large $E$ can also degrade the accuracies of estimate $\tilde{h}_\ell$, due to potential interferences from the other channel taps. Therefore, it is a trade-off between these two different channel types.

The processes (\ref{nl})-(\ref{th1}) are implemented successively for each of the $L$ channel taps. In such a way, solving problem (\ref{prbm}) is transfered to solve a following problem
\bea \label{prbm1} (\tilde{\vec{d}},\tilde{\vec{h}},\tilde{L})\approx\argmin_{\vec{d},\vec{h},L}\sum_{d=0}^{D-1}\left|R[d]-\sum_{i=0}^{L-1}h_i \gamma(d-d_i)\right|^2\!, \eea
which is suboptimal (due to the colored noise in (\ref{R1})), but simplifies the original problem in two aspects: Firstly, the length $D$ of correlations $R[d]$ is in general much shorter then the length $N$ of NPRS $s[k]$, hence, the number of complex multiplications are reduced. Secondly, only one cross-correlation process is needed to compute $R[d]$ in solving (\ref{prbm1}).

To summarize, the SAGE algorithm in the first main step runs iteratively with three sub-steps: First, the ACFs corresponding to the other channel taps are removed from the cross-correlations. Second, $\tilde{h}_\ell$ and $\tilde{d}_\ell$ are updated based on the updated $\tilde{R}_\ell[d]$ successively for all taps. Lastly, the reconstructed signal $\tilde{h}_\ell \gamma(d-\tilde{n}_\ell)$ with refined estimates are then removed from $\tilde{R}[d]$ before moving on to the processes of the next path. 

\begin{table*}[t]
\renewcommand{\arraystretch}{1.5}
\centering
\vspace*{1mm}
\caption{Delay Profile in Number of samples and Relative Power in dB at different sampling-rate.}
\label{tab1}
\begin{tabular}{|c|c|c|c|}
\hline
Channel Type & 30.72 MHz &1.92 MHz& 240 KHz\\ \hhline{|=|=|=|=|} 
        \multirow{2}{4em}{EPA} &[0, 1, 2, 3, 6, 13] samples& [0, 1] samples&0 sample \\ 
   &  [0, -1.0, -2.0, -3.0, -8.0, -17.2, -20.8] dB&  [0.0, -25.7] dB  & 0 dB   \\ \hline
      \multirow{2}{4em}{EVA} &[0, 1, 5, 10, 11, 22, 33, 53, 77] samples& [0, 1, 2, 3, 5] samples&[0, 1] sample\\ 
   & [0, -1.5, -1.4, -3.6, -0.6, -9.1, -7.0, -12.0, -16.9] dB&  [ 0, -2.3, -10.9, -15.9, -20.8] dB  & [0, -23.1] dB  \\ \hline
     \multirow{2}{4em}{ETU} &[0, 2, 4, 6, 7, 15, 49, 71, 154] samples& [0, 1, 3, 4, 10]  samples&[0, 1] sample\\ 
   & [-1.0, -1.0, -1.0, 0, 0, 0, -3.0, -5.0, -7.0] dB&  [0, -6.4, -9.4, -11.4, -13.4] dB  & [0, -14.9] dB  \\ \hline
\end{tabular}
\vspace{-0mm}
\end{table*}

\begin{algorithm}[t!]
        \caption{SAGE based Channel Estimation}\label{alg1}
        \begin{algorithmic}[1]
            \Require Normalized auto-correlation $\gamma(d)$, cross-correlation $R[d]$, tap-length $L$ and SAGE iteration number $M$.
            \State If no initial inputs $\vec{d}$ and $\vec{h}$ available: Sort $|R[d]|$ and set an initial $\tilde{\vec{d}}$ to be the locations of the $L$ largest values, and the channel estimate $\tilde{h}_\ell\!=\!R[\tilde{d}_\ell]$ for all $0\!\leq\!\ell\!<\!L$.
             \For{$\ell=0,1,\cdots,L-1$}
               \State $\tilde{R}[d]=R[d]-\sum\limits_{i=0}^{L-1}\tilde{h}_\ell \gamma(d-\tilde{d}_\ell)$
               \EndFor
                                    \If{$L=1$}
                        \State Break
                   \Else 
                \For{$m=0,1,\cdots,M-1$}
           \For{$\ell=0,1,\cdots,L-1$}
            \State $\tilde{R}[d]=\tilde{R}[d]+\tilde{h}_\ell \gamma(d-\tilde{d}_\ell)$
            \State $\tilde{\sigma}^2\!=\!\left|\frac{1}{D}\sum\limits_{d=0}^{D-1}\tilde{R}[d]\right|^2$
            \State $\tilde{d}_\ell=\argmax\limits_{d}|\tilde{R}[d]|$
            \State $\tilde{h}_\ell=\frac{1}{2E}\sum_{k=-E}^{E}\gamma^{\ast}(k)\tilde{R}_{\ell}[\tilde{d}_\ell+k]$
             \State $\tilde{h}_\ell=\frac{\tilde{h}_\ell}{1+\tilde{\sigma}^2/|\tilde{h}_\ell|^2}$
            \State $\tilde{R}[d]=\tilde{R}[d]-\tilde{h}_\ell \gamma(d-\tilde{d}_\ell)$
            \EndFor
            \EndFor
             \EndIf
            \State Output estimated $\tilde{\vec{h}}$, $\tilde{\vec{d}}$ and $\tilde{\sigma}^2$.
        \end{algorithmic}
    \end{algorithm}
    
         \begin{algorithm}[t!]
        \caption{Proposed ToA estimator for NB-IoT}\label{alg2}
        \begin{algorithmic}[1]
            \Require An initial input $L_0$.
            \State Run Algorithm 1 with $L\!=\!L_0$, and output $\tilde{\vec{h}}$, $\tilde{\vec{d}}$ and  $\tilde{\sigma}^2$.
             \State With input $\tilde{\vec{h}}$ and $\tilde{\vec{d}}$, output refined $\tilde{\vec{h}}$ and $\tilde{\vec{d}}$ after removing invalid paths satisfying (\ref{ref1}). 
                          \State With input $\tilde{\vec{h}}$ and $\tilde{\vec{n}}$, output further refined $\tilde{\vec{h}}$, $\tilde{\vec{d}}$ and $L$ after removing invalid paths satisfying (\ref{ref2}). 
            \While {$L< L_0$}
               \State Set $L_0\!=\!L$.
               \State Repeat Step 1 with initial inputs $\tilde{\vec{d}}$ and $\tilde{\vec{h}}$.
               \State Repeat  Step 2 and Step 3.
               \EndWhile
                  \State If condition (\ref{ref3}) is satisfied, set $L\!=\!1$, $\tilde{d}_0\!=\!d_{\mathrm{peak}}$, and $\tilde{h}_0\!=\!h_{\mathrm{peak}}$; Otherwise, go to the next step.
                           \State Output a final ToA estimate $\tilde{d}_0$.
        \end{algorithmic}
    \end{algorithm}

\subsection{Step 2: Refinement of $L$}
With the estimated $\tilde{\vec{h}}$ and $\tilde{\vec{d}}$, we next propose heuristic methods to detect $L$ and remove invalid channel taps, which comprises two steps and is summarized in Algorithm 2. 

We first consider the case that only one-path is presented, i.e., $L\!=\!1$, then from (\ref{prbm}), the ML estimates \cite{K93} of channel delay and coefficient are
{\setlength\arraycolsep{2pt}\bea \label{dpeak} \tilde{d}_{\mathrm{peak}}&=&\argmax_{d}R[d],  \\
\label{hpeak} \tilde{h}_{\mathrm{peak}}&=&R[\tilde{d}_{\mathrm{peak}}].\eea}
\hspace{-1.4mm}These two estimates plays an key role in finding possible invalid paths.

The first heuristic step is to remove channel taps that are far away from the peak position $\tilde{d}_{\mathrm{peak}}$. That is, if 
\bea \label{ref1} |\tilde{h}_\ell-\tilde{d}_{\mathrm{peak}}|>d_{\mathrm{max}} \eea
is satisfied, $\tilde{h}_\ell$ is classified as an invalid path and removed from $\tilde{\vec{h}}$, and $\tilde{\vec{d}}$ is updated accordingly. Then, $L$ is also updated as the number of remaining taps. The variable $d_{\mathrm{max}}$ is the maximum time-difference that is allowed. The rationale behind this step is that, the first-path should not be too far away from the position where a peak is found, especially at low SNRs. For instance, as seen from Table 1, the largest time-difference between all channel taps for three fading channels are below 10 at sampling rate 1.92 MHz, and $d_{\mathrm{max}}$ can be set to 10 in this case.

The second heuristic step is to recursively remove the channel taps with power less than a fraction of the total channel power. That is, if
\bea  \label{ref2} |\tilde{h}_\ell|^2<\eta_3\bigg(\sum\limits_{0\leq\ell<L}|\tilde{h}_\ell|^2\bigg) , \eea
holds, then the $\ell$-th path is removed from $\tilde{\vec{h}}$ and so is the delay $\tilde{n}_\ell$ in $\tilde{\vec{n}}$, and $L$ is also updated as $L\!-\!1$. Such a process is repeated until the condition (\ref{ref2}) is violated for all remaining taps. The threshold $\eta_3$ is a also pre-defined and yet to be optimized, which can be set to, for instance, 0.1.

\subsection{The Proposed ToA Estimation Method}
Based on the introduced two steps, the proposed ToA estimation method is then summarized in Algorithm 2, which starts with a sufficient large $L$ and run Algorithm 1 to generate initial estimates of $\tilde{\vec{h}}$ and $\tilde{\vec{d}}$. Then, the estimates are refined by removing the channel taps that satisfy conditions (\ref{ref1}) and (\ref{ref2}). If the number of channel taps afterwards is less than the initial value, using the updated $L$ and refined estimates $\tilde{\vec{h}}$, $\tilde{\vec{d}}$ as initializations to run Algorithm 1 again, followed by removing the invalid paths and the refinement of $L$. These two steps run iteratively until $L$ converges. With the last outputs $\tilde{\vec{h}}$, $\tilde{\vec{d}}$ and $\tilde{\sigma}^2$, there is one more step to improving the ToA estimation for a single-path and AWGN channels. That is, if the output noise power $\tilde{\sigma}^2$ is higher than the noise power estimated with $d_{\mathrm{peak}}$ and $h_{\mathrm{peak}}$ in (\ref{dpeak}) and (\ref{hpeak}), i.e.,
\bea \label{ref3} \tilde{\sigma}^2\geq \frac{1}{D}\sum_{d=0}^{D-1}\big|R[d]-h_{\mathrm{peak}}\gamma(d-d_{\mathrm{peak}})\big|^2 \eea
then we claim that there is only one path exist and set the ToA estimate to $d_{\mathrm{peak}}$. This is due to the sub-optimality of the proposed algorithm, which may converge to a local optimum especially when the true condition is that only a single-path is present.

\section{Numerical Results}
In this section, we provide numerical results to show the performance of the proposed ToA estimator. Without loss of generality, we assume a sampling frequency of 1.92 MHz for ToA estimation with a sample time $16T_s$, due to low-end analog devices used in NB-IoT receivers. We make no attempt to optimize the parameters and set $E\!=\!4$ in (\ref{th}), $d_{\mathrm{max}}\!=\!10$ in (\ref{ref1}) and $\eta_3\!=\!0.1$ in (\ref{ref2}). The true value of ToA is set to 50 samples. The number of SAGE iterations in Algorithm 1 is set to $M\!=\!8$ with an initial $L\!=\!5$ for all test cases. For comparison, we also set $\eta_2\!=\!0.5$ in (\ref{threshold2}) for a traditional threshold based ToA estimator as in \cite{Ret15}.

\subsection{PAPR Improvement for False-Alarm Probability}
\begin{figure}[t]
\begin{center}
\scalebox{0.29}{\includegraphics{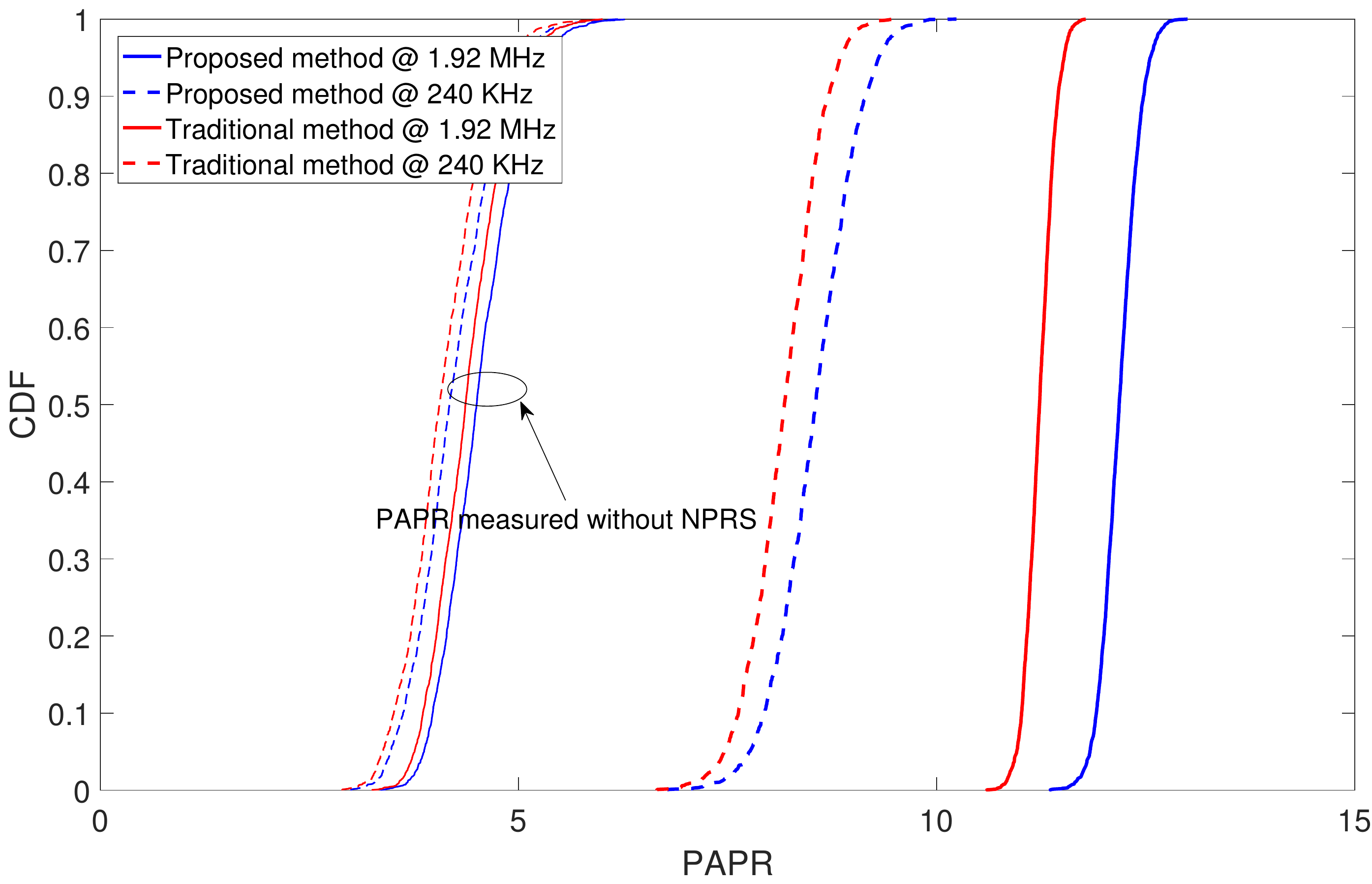}}
\vspace*{-3mm}
\caption{\label{fig4}The PAPR improvements with considering the ACF of NPRS under AWGN channel at SNR  -4 dB.}
\vspace*{-4mm}
\end{center}
\end{figure}

First we show the PAPR improvements with considering the impact of NPRS ACF under AWGN channel. We evaluated at sampling rate 1.92 MHz and 240 KHz, respectively. The PAPR is measured in (\ref{threshold1}) for both methods, while for the proposed method, $R[d]$ in the denominator is replaced by $\tilde{R}_\ell[d]$ in (\ref{tR1}) that removes the ACF centered at the peak position. As can be seen from Fig. \ref{fig4}, when there is no NPRS presented, the CDFs of the PAPR are almost the same for both methods at different sampling rates. However, when NPRS is presented, the CDFs of PAPR are further pushed to right of the proposed method compared to the traditional method. That means, the false-alarm probability of detecting NPRS is improved by removing the impact of the ACF in the proposed method.

\subsection{Convergence of the SAGE Algorithm}
\begin{figure}
\begin{center}
\scalebox{0.29}{\includegraphics{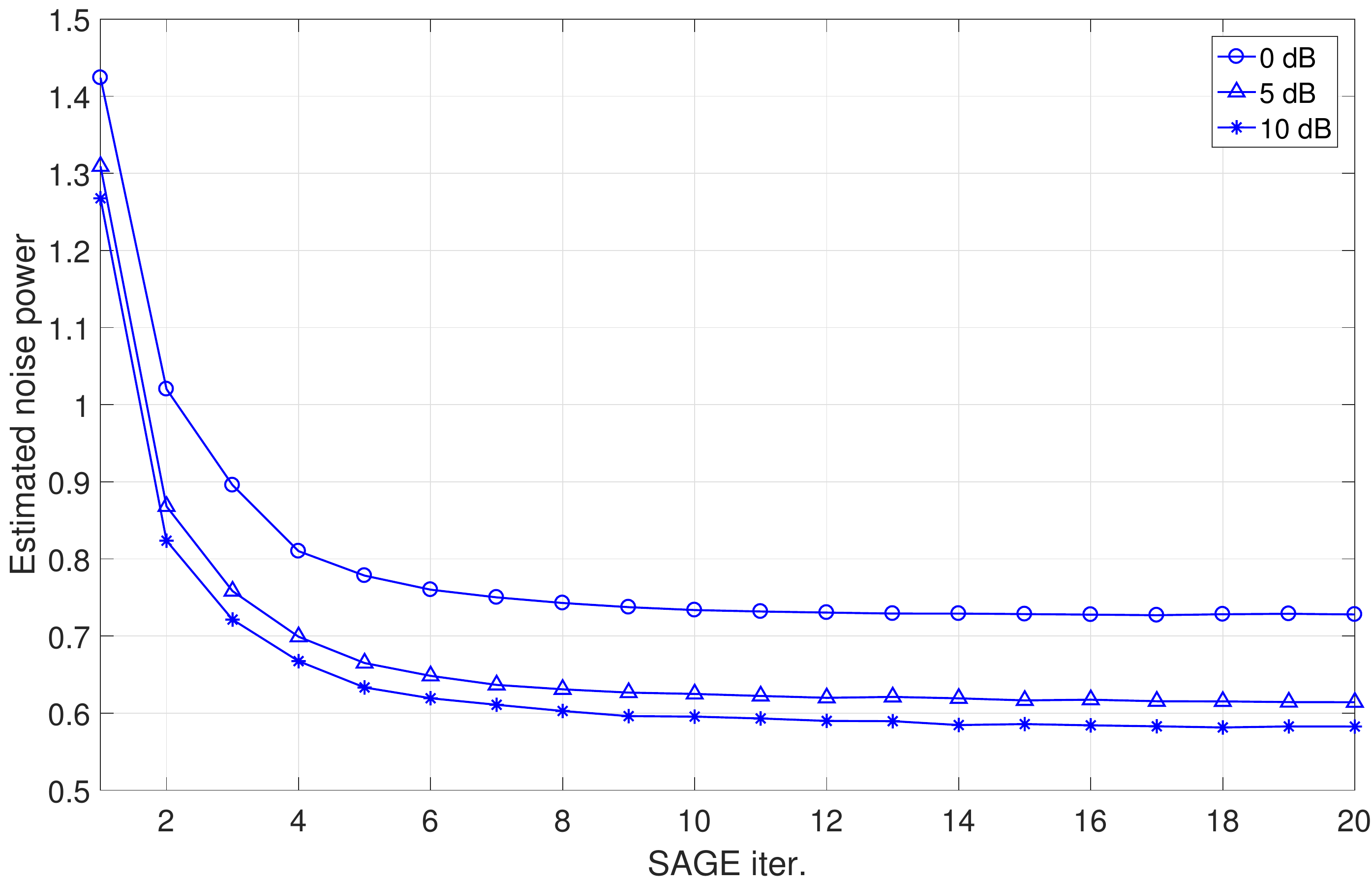}}
\vspace*{-3mm}
\caption{\label{fig5}Convergence of Algorithm 1 in the first running under ETU-3Hz channel. Each curve is averaged over 2000 channel realizations.}
\vspace*{-7mm}
\end{center}
\end{figure}

Next we show the convergence speed of the proposed SAGE Algorithm under ETU-3Hz channel at SNR 5 dB. We show the estimated noise power $\tilde{\sigma}^2$ that are averaged over 2000 channel realizations. As can be seen in Fig. \ref{fig5}, the algorithm converges fast in around 8$\sim$10 SAGE iterations.

\begin{figure}[t]
\begin{center}
\scalebox{0.294}{\includegraphics{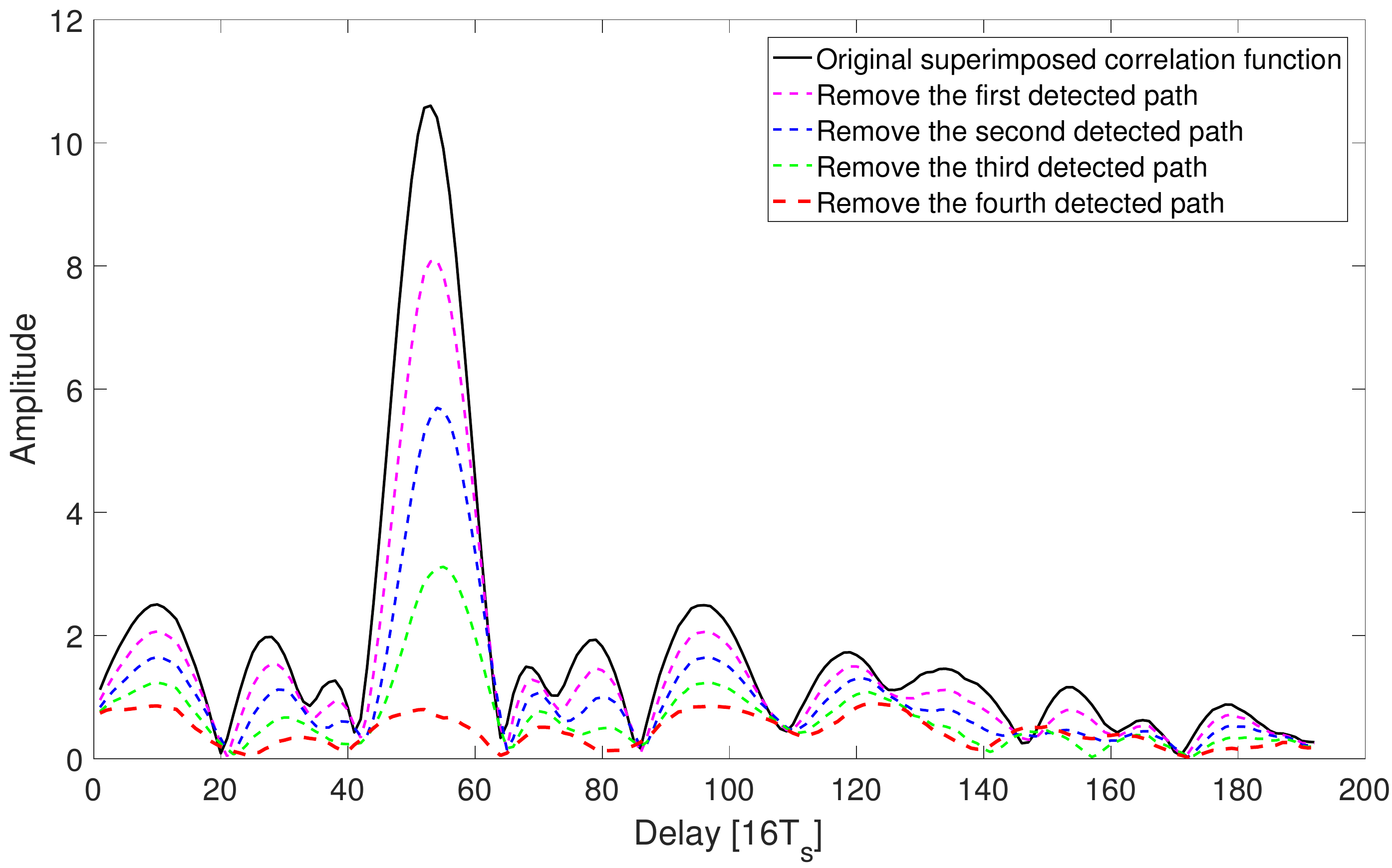}}
\vspace*{-3mm}
\caption{\label{fig6}The amplitude of the correlation $\tilde{R}[d]$ with removing the four detected paths under ETU-3Hz Channel at SNR 5 dB.}
\vspace*{-4mm}
\end{center}
\end{figure}

We also show an example of $\tilde{R}_\ell[d]$ with removing the ACFs corresponding to each of the detected channel path in Fig. \ref{fig6} with the proposed method. The peak of $R[d]$ is attained at $\tilde{d}_{\mathrm{peak}}\!=\!53$ while the true value is 50. With the proposed algorithm, the ToA can be correctly detected with four channel taps detected at delays [50, 51, 53, 54] and estimated channel coefficients $\vec{h}\!=\![0.2882-2.7396i, \,0.0076-2.7314i, \,-0.5047-2.5380i, \,-0.6967-2.3480i]$. In this case, the fifth channel tap is undetected due to its low power. As it can be seen that, with removing each of the four detected path component successively, the amplitude of $\tilde{R}_\ell[d]$ decreases and acts like a noise floor at last.

\subsection{ToA Detection Performance}
Last we show the ToA detection performance with the proposed ToA estimation method under different channels. We measure both the exact detection probability, and the detection probability that the estimate errors are less than 3 samples. 

\begin{figure}
\begin{center}
\scalebox{0.29}{\includegraphics{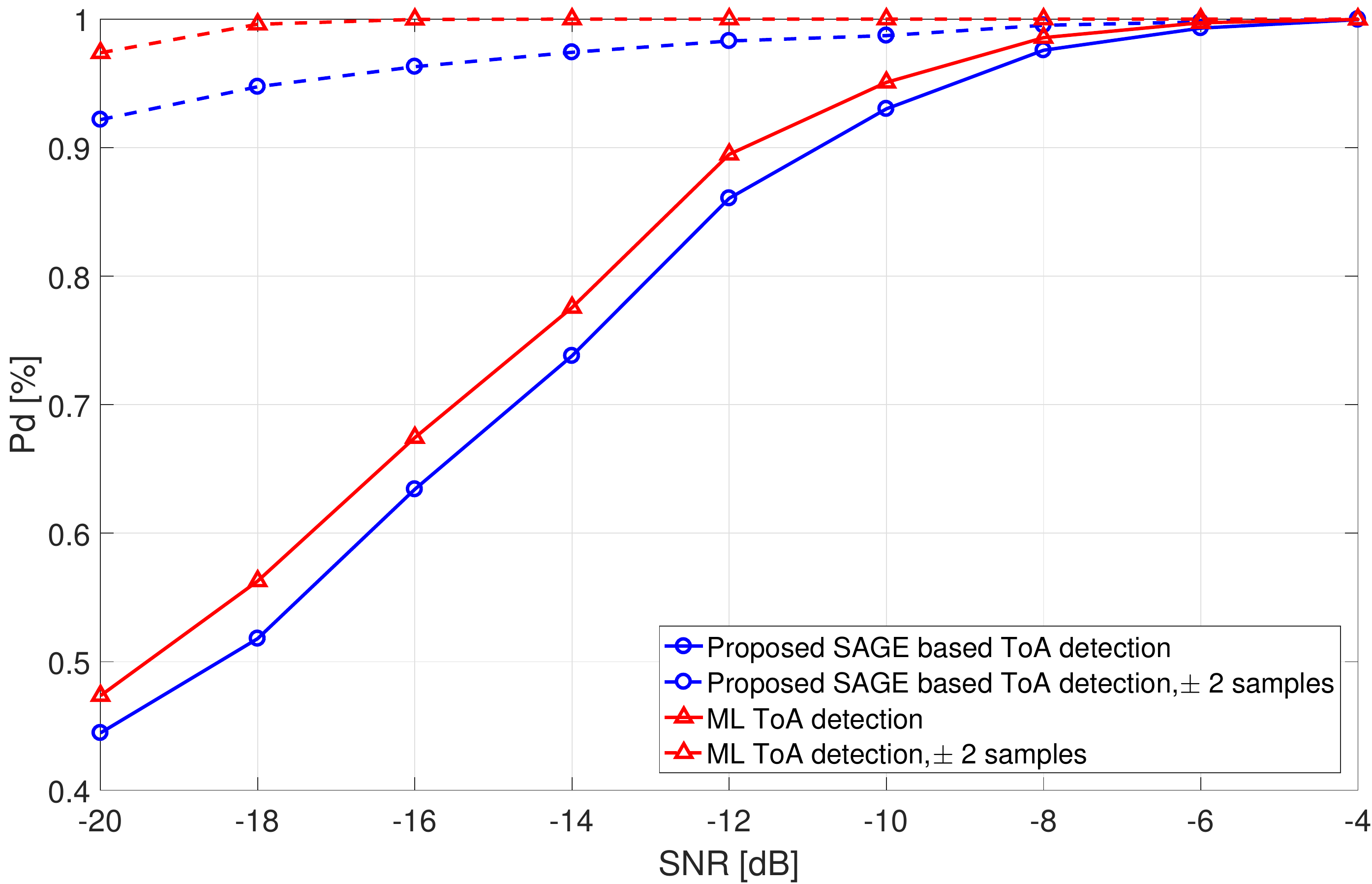}}
\vspace*{-3mm}
\caption{\label{fig7}ToA detection performance under ETU-3Hz channel.}
\vspace*{-7mm}
\end{center}
\end{figure}

In Fig. \ref{fig7}, the detection probabilities of the proposed method are compared to the ML estimator under AWGN channel, where the ML estimator utilizes the priori information that only one channel tap exists and the ML estimates are in (\ref{dpeak}) and (\ref{hpeak}), while the proposed method does not use such a priori information. As can be seen, the proposed method performs close to the ML in both cases.

\begin{figure}[t]
\begin{center}
\scalebox{0.29}{\includegraphics{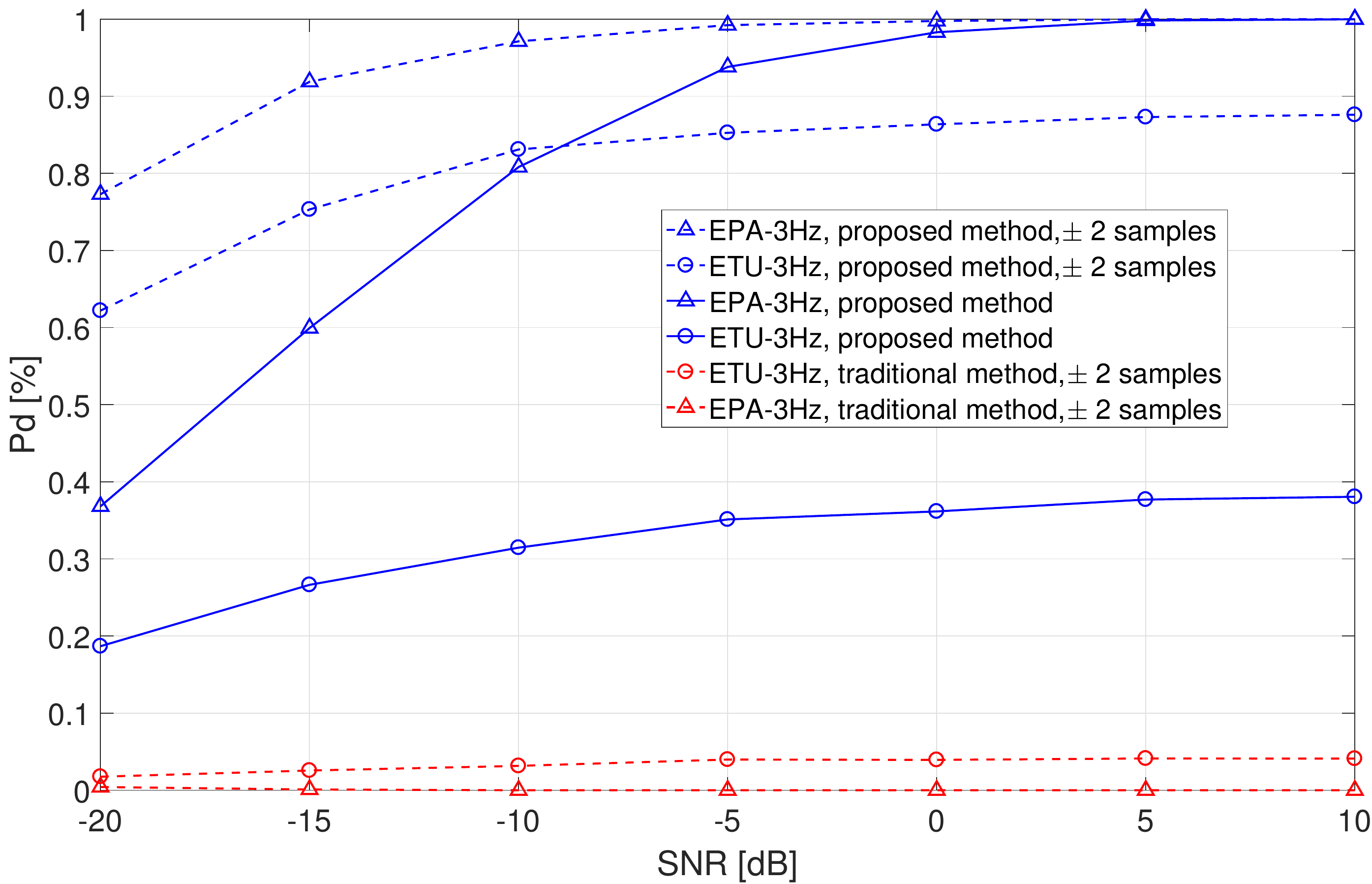}}
\vspace*{-3mm}
\caption{\label{fig8}ToA detection performance under EPA-3Hz and ETU-3Hz Channel.}
\vspace*{-7mm}
\end{center}
\end{figure}

In Fig. \ref{fig8}, the detection probabilities of the proposed method are compared to the traditional estimator under fading channels. As can be seen, the traditional threshold based method performs quite poor under both EPA-3Hz and ETU-3Hz channels even with tolerable estimate errors less than 3 sample. With tolerable estimate errors less than 3 sample, even at SNR -15dB, the detection probability using the proposed ToA estimation method is above 75\% and 90\% under ETU-3Hz and EPA-3Hz channels, respectively. The performance under EPA-3Hz channel is better due to the fact that, as seen from Table 1, at sampling rate 1.92 MHz, EPA channel degrades to a single path channel (since the power of the second path is negligible), while ETU-3Hz channel comprises 5 taps that make it is more difficult for ToA estimation.

\section{Summary}
We have considered time-of-arrival (ToA) estimation in narrowband Internet-of-Things (NB-IoT) systems and proposed a low-complexity space-alternating generalized expectation-maximization (SAGE) based method for joint estimating the number of channel taps, the channel coefficients and the delays. Due to a limited bandwidth 180 KHz and the number of NPRS symbols used, the time-domain auto-correlation function (ACF) of the NB positioning reference signal (NPRS) is not perfect and has a wide lobe. By taking this impairment into account, in a first step of the proposed ToA estimation method, the SAGE algorithm is utilized to decompose the superimposed correlations corresponding to different channel taps and generate estimates for channel coefficients and the corresponding delays. Then in a second step, the estimated channel taps and number of taps are refined by removing invalid paths. These two steps run iteratively until the number of estimated channel taps is converged. We show through simulation results that, the proposed ToA estimation method performs close to the maximum likelihood (ML) estimator for a single-path channel, and outperforms a traditional threshold based ToA estimator without considering the impacts of multiple-path channel and the ACF of NPRS in NB-IoT systems.

\end{document}